\documentclass[sigconf]{acmart}

\renewcommand\footnotetextcopyrightpermission[1]{}
\usepackage{tikz}
\usepackage[all]{xy}
\usetikzlibrary{arrows,shapes}

\AtBeginDocument{%
  \providecommand\BibTeX{{%
    \normalfont B\kern-0.5em{\scshape i\kern-0.25em b}\kern-0.8em\TeX}}}

\setcopyright{acmcopyright}
\copyrightyear{2020}
\acmYear{2020}
\acmDOI{}

\acmConference[]{}
\acmBooktitle{2020}
\acmPrice{15.00}
\acmISBN{978-1-4503-9999-9/18/06}

\begin{document}

%%
%% The "title" command has an optional parameter,
%% allowing the author to define a "short title" to be used in page headers.
\title{Path homology as a stronger analogue of cyclomatic complexity}

%%
%% The "author" command and its associated commands are used to define
%% the authors and their affiliations.
%% Of note is the shared affiliation of the first two authors, and the
%% "authornote" and "authornotemark" commands
%% used to denote shared contribution to the research.
\author{Steve Huntsman}
\email{steve.huntsman@baesystems.com}
\affiliation{%
  \institution{BAE Systems FAST Labs}
  \streetaddress{4301 North Fairfax Drive, Suite 800}
  \city{Arlington}
  \state{Virginia}
  \postcode{22202}
}

%%
%% By default, the full list of authors will be used in the page
%% headers. Often, this list is too long, and will overlap
%% other information printed in the page headers. This command allows
%% the author to define a more concise list
%% of authors' names for this purpose.
\renewcommand{\shortauthors}{Huntsman}

%%
%% The abstract is a short summary of the work to be presented in the
%% article.
\begin{abstract}
	{\bf Background.} Cyclomatic complexity is an incompletely specified but mathematically principled software metric that can be usefully applied to both source and binary code.

	{\bf Aims.} We consider the application of path homology as a stronger analogue of cyclomatic complexity. 
	
	{\bf Method.} We have implemented an algorithm to compute path homology in arbitrary dimension and applied it to several classes of relevant flow graphs, including randomly generated flow graphs representing structured and unstructured control flow. We also compared path homology and cyclomatic complexity on a set of disassembled binaries obtained from the \texttt{grep} utility.
	
	{\bf Results.} There exist control flow graphs realizable at the assembly level with nontrivial path homology in arbitrary dimension. We exhibit several classes of examples in this vein while also experimentally demonstrating that path homology gives identicial results to cyclomatic complexity for at least one detailed notion of structured control flow. We also experimentally demonstrate that the two notions differ on disassembled binaries, and we highlight an example of extreme disagreement.
	
	{\bf Conclusions.} Path homology empirically generalizes cyclomatic complexity for an elementary notion of structured code and appears to identify more structurally relevant features of control flow in general. Path homology therefore has the potential to substantially improve upon cyclomatic complexity.
% Cyclomatic complexity is an incompletely specified but mathematically principled software metric that can be usefully applied to both source and binary code. We consider the application of path homology as a more powerful analogue of cyclomatic complexity. There exist control flow graphs realizable at the assembly level with nontrivial path homology in arbitrary dimension. We exhibit several classes of examples in this vein while also experimentally demonstrating that path homology gives identicial results to cyclomatic complexity for at least one detailed notion of structured control flow. Thus path homology empirically generalizes cyclomatic complexity, and has the potential to substantially improve upon it.
\end{abstract}

%%
%% The code below is generated by the tool at http://dl.acm.org/ccs.cfm.
%% Please copy and paste the code instead of the example below.
%%
\begin{CCSXML}
<ccs2012>
<concept>
<concept_id>10002944.10011123.10011124</concept_id>
<concept_desc>General and reference~Metrics</concept_desc>
<concept_significance>500</concept_significance>
</concept>
<concept>
<concept_id>10002950.10003741.10003742.10003744</concept_id>
<concept_desc>Mathematics of computing~Algebraic topology</concept_desc>
<concept_significance>300</concept_significance>
</concept>
<concept>
<concept_id>10002950.10003624.10003633.10003640</concept_id>
<concept_desc>Mathematics of computing~Paths and connectivity problems</concept_desc>
<concept_significance>300</concept_significance>
</concept>
</ccs2012>
\end{CCSXML}

\ccsdesc[500]{General and reference~Metrics}
\ccsdesc[300]{Mathematics of computing~Algebraic topology}
\ccsdesc[300]{Mathematics of computing~Paths and connectivity problems}

%%
%% Keywords. The author(s) should pick words that accurately describe
%% the work being presented. Separate the keywords with commas.
\keywords{software metric, cyclomatic complexity, path homology}

%%% A "teaser" image appears between the author and affiliation
%%% information and the body of the document, and typically spans the
%%% page.
%\begin{teaserfigure}
%  \includegraphics[width=\textwidth]{sampleteaser}
%  \caption{Seattle Mariners at Spring Training, 2010.}
%  \Description{Enjoying the baseball game from the third-base
%  seats. Ichiro Suzuki preparing to bat.}
%  \label{fig:teaser}
%\end{teaserfigure}

%%
%% This command processes the author and affiliation and title
%% information and builds the first part of the formatted document.
\maketitle

\section{\label{sec:Introduction}Introduction}

An archetypal software metric is the \emph{cyclomatic complexity} of the control flow graph of a computer program \cite{McCabe1976}, i.e., the number of linearly indepdendent paths through the control flow. A more explicitly topological formulation motivating the present paper is that cyclomatic complexity is the first Betti number of the undirected graph regarded as a simplicial 1-complex, or in less mathematical language, the number of ``holes'' in the control flow graph.

Ironically in light of the preceding technical characterizations, cyclomatic complexity is \cite{Ebert2016} 
\begin{quote}
shunned for the most part by academia for certain theoretical weaknesses \dots paradoxically, industry uses it extensively \dots [it] has great benefit for projects to predict software components that likely have a high defect rate or that might be difficult to test and maintain. It's of even more value having a simple indicator that can provide constructive guidance on how to improve the code's quality. 
\end{quote}

Meanwhile, there are no widely agreed-upon alternatives to cyclomatic complexity, much less any with a comparably mathematical underpinning that permits principled adaptation and generalization \cite{Ajami2019}. Moreover, at least some of the criticism directed at cyclomatic complexity stems from ambiguity of control flow representations of source code that disappear at the level of disassembled binary code, where every control flow operation is a jump of some kind, and where cyclomatic complexity can usefully guide testing and reverse engineering efforts such as fuzzing \cite{Iozzo2010,Duran2011}. Even at the source code level, cyclomatic complexity and other software metrics can contribute substantially to the identification of fault-prone or vulnerable code \cite{Alves2016,Medeiros2017,Du2019}. In short, cyclomatic complexity is an incomplete but principled software metric with useful applications to both source and binary code. 

Here, we give theoretical and experimental evidence (but not yet a mathematical proof) that the recently developed theory of \emph{path homology} \cite{Grigoryan2012,Grigoryan2014,Grigoryan2014b,Grigoryan2015,Grigoryan2017,Grigoryan2018,Grigoryan2018b} gives a metric that not only broadly coincides with cyclomatic complexity in dimension one for an elementary notion of structured control flow, but is also more powerful than cyclomatic complexity in determining the presence of control flow regions that behave like holes in arbitrarily high dimension, e.g., ``voids'' in dimension two. While these regions can be more difficult to interpret geometrically, the analogy is mathematically precise, and path homology also offers a mechanism towards principled feature construction in efforts that seek to couple machine learning with software engineering (for which see, e.g. \cite{Allamanis2018}).

In particular, path homology both generalizes the simplicial homology theory underlying cyclomatic complexity (though not cyclomatic complexity \emph{per se}) and directly applies to digraphs. Whereas the simplicial Betti numbers of a 1-complex equal the cyclomatic complexity in dimension 1 and are trivial in all higher dimensions (see \S \ref{sec:CyclomaticComplexity}), the path homology of a digraph is a richer topological invariant whose Betti numbers can take any values in every dimension. It is therefore natural to consider its application as a more powerful analogue of cyclomatic complexity that is especially well suited for reverse engineering and vulnerability testing efforts.

\section{\label{sec:Homology}Homology}

For background on algebraic topology, including the theory of simplicial homology that often introduces the subject and shares many similarities with path homology, see \cite{Ghrist2014,Hatcher2002,Kozlov2008}.

\begin{figure}[htbp]
\includegraphics[trim = 60mm 110mm 60mm 115mm, clip, width=.9\columnwidth,keepaspectratio]{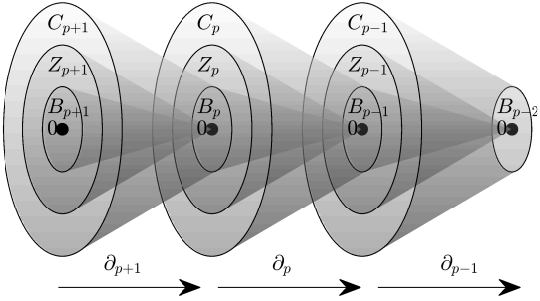}% Here is how to import pix
\caption{ \label{fig:ChainComplex} Schematic picture of a chain complex.
} 
\end{figure}

A \emph{chain complex} over a field $\mathbb{F}$ is a pair of sequences (indexed by a ``dimension'' $p \in \mathbb{N}$) of $\mathbb{F}$-vector spaces $C_p$ and linear \emph{boundary operators} $\partial_p : C_p \rightarrow C_{p-1}$ such that $\partial_{p-1} \circ \partial_p \equiv 0$. We can visualize such a structure as in Figure \ref{fig:ChainComplex}, and write it as
\begin{equation}
\label{eq:chainComplex}
\dots C_{p+1} \overset{\partial_{p+1}}{\longrightarrow} C_p \overset{\partial_p}{\longrightarrow} C_{p-1} \overset{\partial_{p-1}}{\longrightarrow} \dots \overset{\partial_1}{\longrightarrow} C_0 \overset{\partial_0}{\longrightarrow} 0.
\end{equation}
Writing $Z_p := \text{ker } \partial_p$ and $B_p := \text{im } \partial_{p+1}$, the \emph{homology} of \eqref{eq:chainComplex} is 
\begin{equation}
H_p := Z_p/B_p.
\end{equation}
The \emph{Betti numbers} are $\beta_p := \text{dim } H_p = \text{dim } Z_p - \text{dim } B_p$.

% ChainComplexPic.m

%In practice, it is frequently convenient for technical reasons to (and we do) work with the \emph{reduced} homology $\tilde H_p$. This has the minor effect $\tilde H_0 \oplus \mathbb{F} \cong H_0$, while $\tilde H_p \cong H_p$ for $p > 0$. Similarly, and using an obvious notational device, $\tilde \beta_p = \beta_p - \delta_{p0}$, where $\delta_{jk} = 1$ iff $j = k$ and $\delta_{jk} = 0$ otherwise.

In practice, it is frequently convenient for technical reasons to (and we do) replace the chain complex \eqref{eq:chainComplex} with its \emph{reduction}
\begin{equation}
\label{eq:reduction}
\dots C_{p+1} \overset{\partial_{p+1}}{\longrightarrow} C_p \overset{\partial_p}{\longrightarrow} C_{p-1} \overset{\partial_{p-1}}{\longrightarrow} \dots \overset{\partial_1}{\longrightarrow} C_0 \overset{\tilde \partial_0}{\longrightarrow} \mathbb{F} \longrightarrow 0
\end{equation}
which (using an obvious notational device and assuming the original chain complex is nondegenerate) has the minor effect $\tilde H_0 \oplus \mathbb{F} \cong H_0$, while $\tilde H_p \cong H_p$ for $p > 0$. Similarly, $\tilde \beta_p = \beta_p - \delta_{p0}$, where $\delta_{jk} = 1$ iff $j = k$ and $\delta_{jk} = 0$ otherwise.

\subsection{\label{sec:Simplicial}Simplicial homology}

Following \cite{Kozlov2008}, we now proceed to sketch the archetypal notion of \emph{simplicial homology} that underlies cyclomatic complexity. An \emph{abstract simplicial complex} (ASC) is a family $\Delta$ of finite subsets $\{v_0, \dots, v_p\}$ (called \emph{$p$-simplices}) of a set $V$ of \emph{vertices} such that if $X \in \Delta$ and $\varnothing \ne Y \subseteq X$, then $Y \in \Delta$.
\footnote{
In other words, an ASC is a hypergraph with all sub-hyperedges. 
}

Given an ASC $\Delta$, let $C_p$ be the $\mathbb{F}$-vector space generated by basis elements $e_{(v_0,\dots,v_p)}$ corresponding to \emph{oriented simplices} of \emph{dimension} $p$ in $\Delta$. This essentially means that if $\sigma$ is a permutation acting on $(v_0,\dots,v_p)$, then $e_{(v_0,\dots,v_p)} = (-1)^\sigma e_{(v_{\sigma(0)},\dots,v_{\sigma(p)})}$.
\footnote{
Thus for example $e_{(v_0,v_1,v_2)} = -e_{(v_0,v_2,v_1)}$.
}
\footnote{
Note that an order on $V$ induces an order on each simplex in $\Delta$, and in turn this induces an orientation.
}

The simplicial boundary operator $\partial_p$ is now defined to be the linear map acting on basis elements as
\begin{equation}
\label{eq:simplicial}
\partial_p e_{( v_0,\dots,v_p )} = \sum_{j=0}^p (-1)^j e_{ \nabla_j ( v_0,\dots,v_p )}
\end{equation}
where $\nabla_j$ deletes the $j$th entry of a tuple. It turns out that this construction yields a \emph{bona fide} chain complex. Moreover, the simplicial Betti numbers measure the number of voids of a given dimension in a geometric realization of an ASC.
\footnote{
Here, 0-dimensional voids amount to connected components. 
}

For example, the boundary of a 2-simplex or ``triangle'' is
\begin{equation}
\partial_2 e_{(1,2,3)} = e_{(2,3)} - e_{(1,3)} + e_{(1,2)} = e_{(1,2)} + e_{(2,3)} + e_{(3,1)} \nonumber
\end{equation}
and its boundary in turn is
\begin{equation}
\partial_1 \left ( e_{(1,2)} + e_{(2,3)} + e_{(3,1)} \right ) = 0. \nonumber
\end{equation}
Thus the homology of the boundary of a triangle has $\beta_p = \delta_{p1}$: there is a single void in dimension 1, and none in other dimensions.

\subsection{\label{sec:CyclomaticComplexity}Cyclomatic complexity}

The \emph{cyclomatic number} or \emph{cyclomatic complexity} \cite{McCabe1976} of an undirected connected graph $G = (V,E)$ is $\nu(G) := |E|-|V|+1$: a result dating to Euler (for a modern treatment, see, e.g. \cite{Bollobas2002}) is that this equals the dimension of the so-called cycle space of $G$. Meanwhile, the cyclomatic number also equals the first (simplicial) Betti number of the abstract simplicial complex whose $2$-simplices correspond to edges \cite{Serre1980}, and all Betti numbers in dimension $>1$ are identically zero.

\subsection{\label{sec:PathHomology}Path homology}

Our sketch of path homology mostly follows \cite{Grigoryan2012,Chowdhury2018}, with some small changes to notation and terminology that should be easily handled by the interested reader.

Let $D = (V,A)$ be a loopless digraph and let $p \in \mathbb{Z}_+$. The set $\mathcal{A}_p(D)$ of \emph{allowed $p$-paths} is
\begin{equation}
\label{eq:allowed}
\{(v_0,\dots,v_p) \in V^{p+1} : (v_{j-1},v_j) \in A, 1 \le j \le p\}.
\end{equation}
By convention, we set $\mathcal{A}_{0} := V$, $V^0 \equiv \mathcal{A}_{-1} := \{0\}$ and $V^{-1} \equiv \mathcal{A}_{-2} := \varnothing$. For a field $\mathbb{F}$ (in practice, we take $\mathbb{F} = \mathbb{R}$)
\footnote{
Path homology can be defined over rings as well, and M. Yutin has exhibited digraphs that illustrate the nontriviality of this definition, i.e. \emph{torsion}. It turns out that one such digraph on just six vertices can even be realized as an assembly-level control flow graph, though not a flow graph in the sense of \S \ref{sec:FlowGraphs}.
}
and a finite set $X$, let $\mathbb{F}^X \cong \mathbb{F}^{|X|}$ be the free $\mathbb{F}$-vector space on $X$, with the convention $\mathbb{F}^\varnothing := \{0\}$. The \emph{non-regular boundary operator} $\partial_{[p]} : \mathbb{F}^{V^{p+1}} \rightarrow \mathbb{F}^{V^p}$ is the linear map acting on the standard basis as
\begin{equation}
\label{eq:preboundary}
\partial_{[p]} e_{(v_0,\dots,v_p)} = \sum_{j=0}^p (-1)^j e_{\nabla_j (v_0,\dots,v_p )}.
\end{equation}
A straightforward calculation shows that $\partial_{[p-1]} \circ \partial_{[p]} \equiv 0$, so $(\mathbb{F}^{V^{p+1}},\partial_{[p]})$ is a chain complex. However, we are not concerned with the homology of this chain complex, but of a derived one.

Towards this end, set
\begin{equation}
\label{eq:invariant}
\Omega_p := \left \{ \omega \in \mathbb{F}^{\mathcal{A}_{p}} : \partial_{[p]} \omega \in \mathbb{F}^{\mathcal{A}_{p-1}} \right \},
\end{equation}
$\Omega_{-1} := \mathbb{F}^{\{0\}} \cong \mathbb{F}$, and $\Omega_{-2} := \mathbb{F}^{\varnothing} = \{0\}$. If $\omega \in \Omega_p$, then automatically $\partial_{[p]} \omega \in \mathbb{F}^{\mathcal{A}_{p-1}}$, so $\partial_{[p-1]} \partial_{[p]} \omega = 0 \in \mathbb{F}^{\mathcal{A}_{p-2}}$. Therefore, $\partial_{[p]} \omega \in \Omega_{p-1}$. We thus get a chain complex $(\Omega_p,\partial_p)$ called the \emph{(non-regular) path complex} of $D$, where $\partial_p := \partial_{[p]}|_{\Omega_p}$.
\footnote{
The implied \emph{regular path complex} amounts to enforcing a condition that prevents a directed 2-cycle from having nontrivial 1-homology. While \cite{Grigoryan2012} advocates using regular path homology, in our view non-regular path homology is simpler, more likely useful in applications such as the present one, and exhbits richer phenomenology.
}
The corresponding homology is the \emph{(non-regular) path homology} of $D$. %We implemented an algorithm to compute this in MATLAB and applied it in the present paper; a more performant implementation building on ours is available at {\bf[REFERENCE OMITTED FOR DOUBLE-BLINDING]}.

\subsubsection{\label{sec:Suspension}Path homologies of suspensions}

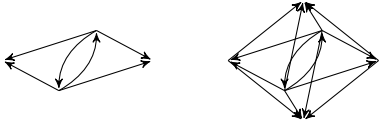
\begin{figure}
	\centering
	\begin{tikzpicture}[every node/.style={inner sep=0,outer sep=0,scale=0.75},->,>=stealth',shorten >=1pt]
	\coordinate (v1) at (.25,.4);
	\coordinate (v2) at (-.25,-.4);
	\coordinate (vA) at (-1,0);
	\coordinate (vB) at (1,0);
	\draw (v1) to [out=-150,in=90,looseness=1] (v2);
	\draw (v1) to (vA);		
	\draw (v1) to (vB);		
	\draw (v2) to [out=30,in=-90,looseness=1] (v1);
	\draw (v2) to (vA);		
	\draw (v2) to (vB);	
	\coordinate (w1) at (3.25,.4);
	\coordinate (w2) at (2.75,-.4);
	\coordinate (w3) at (2,0);
	\coordinate (w4) at (4,0);
	\coordinate (w5) at (3,.8);
	\coordinate (w6) at (3,-.8);
	\draw (w1) to [out=-150,in=90,looseness=1] (w2);
	\draw (w1) to (w3);		
	\draw (w1) to (w4);		
	\draw (w2) to [out=30,in=-90,looseness=1] (w1);
	\draw (w2) to (w3);		
	\draw (w2) to (w4);	
	\foreach \from/\to in {
		w1/w5, w2/w5, w3/w5, w4/w5, w1/w6, w2/w6, w3/w6, w4/w6}
	\draw (\from) to (\to);	
	\end{tikzpicture}
	\caption{ \label{fig:suspensions} These digraphs respectively have $\tilde \beta_2 = 1$ and $\tilde \beta_3 = 1$.}
\end{figure}

Even very simple digraphs can have nontrivial path homology over $\mathbb{R}$ in dimension $\ge 1$. For instance, the digraphs in the left and right panels of Figure \ref{fig:suspensions} respectively have $\tilde \beta_2 = 1$ and $\tilde \beta_3 = 1$. More generally, the $k$-fold ``suspension'' of a directed 2-cycle has $\tilde \beta_{k+1} = 1$ with all other path homologies trivial. The idea is that the $(k-1)$th suspension corresponds to a $k$-dimensional sphere, and the $k$th suspension treats this as the equator of a $(k+1)$-dimensional sphere obtained by adding arcs from each vertex to two new poles. In particular, the digraph in the left panel of Figure \ref{fig:suspensions} corresponds to gluing two cones together, where each cone is obtained in turn by gluing two triangles (2-simplices) together. 

Although these examples cannot be realized in any reasonable notion of control flow, in \S \ref{sec:FlowGraphs} we will see many other digraphs that can be and that have nontrivial path homology in dimension $>1$.

\section{\label{sec:FlowGraphs}Flow graphs}

Following \cite{Huntsman2019} (in this section only), a \emph{flow graph} is a digraph with exactly one source (i.e., a vertex with indegree 0) and exactly one target (i.e., a vertex with outdegree 0) such that there is a unique (entry) arc from the source and a unique (exit) arc to the target, and such that identifying the source of the entry arc with the target of the exit arc yields a strongly connected digraph.
\footnote{
Unlike in \cite{Huntsman2019}, here we disallow loops (i.e., 1-cycles) in digraphs, but the ramifications of this difference are straightforward. In particular, loops in control flow at the assembly level can be transformed into 2-cycles through the introduction of unconditional jumps to/from ``distant'' memory addresses in a way that does little violence to the actual binary code, and no violence at all to its semantics.
}
(We do not require the entry and exit arcs to be distinct, e.g., if the flow graph has only two vertices.) 

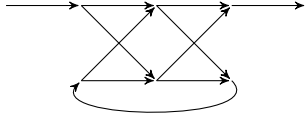
\begin{figure}
	\centering
	\begin{tikzpicture}[every node/.style={inner sep=0,outer sep=0,scale=0.75},->,>=stealth',shorten >=1pt]
	\coordinate (v1) at (0,0);
	\coordinate (v2) at (1,0);
	\coordinate (v3) at (1,-1);
	\coordinate (v4) at (2,0);
	\coordinate (v5) at (2,-1);
	\coordinate (v6) at (3,0);
	\coordinate (v7) at (3,-1);
	\coordinate (v8) at (4,0);
	\foreach \from/\to in {
		v1/v2, v2/v4, v2/v5, v3/v4, v3/v5, v4/v6, v4/v7, v5/v6, v5/v7, v6/v8}
	\draw (\from) to (\to);	
	\draw (v7) to [out=-45,in=-135,looseness=1] (v3);
	\end{tikzpicture}
	\caption{ \label{fig:nontrivialFlow} This digraph can be realized as the assembly-level control flow graph of a program; it has $\tilde \beta_\bullet = (0,1,1,0,\dots)$.}
\end{figure}

The example of Figure \ref{fig:nontrivialFlow} was constructed using the following theorem of \cite{Chowdhury2019}.

\textsc{Theorem.} For $L, n_1,\dots,n_L \in \mathbb{Z}_+$, define the digraph $K^\rightarrow_{n_1,\dots,n_L} := \left ( [N], A_{n_1,\dots,n_L} \right )$ where $N := \sum_{\ell = 1}^L n_\ell$, $[N] := \{1,\dots,N\}$, $A_{n_1,\dots,n_L} := \bigcup_{\ell = 1}^{L-1} A_{n_\ell,n_{\ell+1}}$, and
\begin{equation}
A_{n_\ell,n_{\ell+1}} := \left ( [n_\ell] + \sum_{k = 1}^{\ell-1} n_k \right ) \times \left ( [n_{\ell+1}] + \sum_{k = 1}^{\ell} n_k \right ). \nonumber
\end{equation}
Then
\begin{equation}
\label{eq:NNBetti}
\tilde \beta_{p} \left ( K^\rightarrow_{n_1,\dots,n_L} \right ) = \delta_{p,L-1} \prod_{\ell = 1}^L (n_\ell - 1). \quad \Box
\end{equation}

The proof for a variant of the flow graph above where the bottom arc is replaced by two consecutive arcs follows from the theorem above along with Theorems 5.1 and 5.7 of \cite{Grigoryan2012} (which hold in the non-regular context), so we shall not belabor the exact variant shown; in any case, we can just compute the homology using our algorithm. Furthermore, adding ``layers'' to this example shows how we can produce valid flow graphs realizable as assembly-level control flow graphs and that have nontrivial path homology in arbitrarily high dimension. It is worth noting that control flow graphs similar to those in Figures \ref{fig:nontrivialFlow} and \ref{fig:intel_cfg} can be realized by instantiating the Viterbi algorithm that is ubiquitous in error correction and the analysis of hidden Markov models \cite{MacKay2003}.

\subsection{\label{sec:2FG}2-flow graphs}

We could exhibit nontrivial path homology in dimension $>1$ for many other flow graphs, but we focus here on a highly restricted class of flow graphs to demonstrate that interesting behavior still occurs. The set of flow graphs in which the outdegrees of vertices are all $\le 2$ models control flow at the assembly level in typical architectures: the program counter either advances linearly through memory addresses as operations are executed, or it jumps to a new memory address based on the truth value of a Boolean predicate \cite{Cooper2012,Muchnick1997}. 

In practice, many vertices in assembly-level control flow graphs have outdegree 1, but indegree $> 1$ owing to the presence of inbound jumps.
\footnote{
Such inbound jumps may land in the middle of what would otherwise be a basic block (i.e., a sequence of code statements without any [nondegenerate] control flow), which results in vertices with indegree and outdegree both equal to 1: in fact, a first attempt to produce Figure \ref{fig:intel_cfg} exhibited this behavior.
}
More generally, jumps into or out of an otherwise ``structured'' control flow motif that can be interpreted as an \texttt{if}, a \texttt{while}, or a \texttt{repeat} construct are called ``unstructured'' \texttt{goto}s in the programming literature \cite{Dijkstra1968,McCabe1976,Sennesh2016}. These can be eliminated through control flow restructuring \cite{Bohm1966,Harel1980,Zhang2004}, so that the control flow is ``structured'' in the sense that it can be interpreted as arising from a combination of \texttt{if}, \texttt{while}, and \texttt{repeat}-type structures (though at the assembly level, every control flow operation is instantiated as a jump--i.e., a \texttt{goto}--of some sort, however it may be morally interpreted). 

Rather than precisely define the notion of a structured flow graph \emph{per se}, we consider the simpler, related notion of a \emph{2-flow graph} (2FG). A 2FG is a flow graph with a \emph{source vertex} with outdegree 1, a \emph{target vertex} with outdegree 0, a single vertex adjacent to the target with outdegree 1, and such that all other vertices have outdegree 2. 

Suppose we construct all digraphs on $N$ vertices with outdegree identically 2, except for a single vertex with outdegree 0.
\footnote{
The number of such digraphs follows the sequence at \url{http://oeis.org/A003286}. That is, starting from $N = 3$, there are $1, 7, 66, 916, 16816, \dots$ such digraphs.
}
If such a digraph has a vertex $a$ with nonzero outdegree and such that adding an arc to $a$ from the unique vertex $z$ with outdegree 0 results in a strongly connected digraph, then we call the digraph a \emph{2FG progenitor at $(a,z)$}. Note that a digraph can be a 2FG progenitor at $(a,z)$ and at $(a',z)$ for $a \ne a'$: indeed, the only way such a situation can be avoided is if $a$ has indegree 0, in which case the strong connectivity requirement ensures that $a$ is unique. Given a 2FG progenitor at $(a,z)$, we can construct a 2FG by adding a vertex $s$ and the arc $(s,a)$ and possibly also a vertex $t$ and the arc $(z,t)$. Conversely, removing the entry arc, and possibly also the exit arc, from a 2FG yields a 2FG progenitor, so every 2FG can be constructed from a 2FG progenitor.

\begin{figure}[htbp]
\includegraphics[trim = 50mm 115mm 50mm 115mm, clip, width=\columnwidth,keepaspectratio]{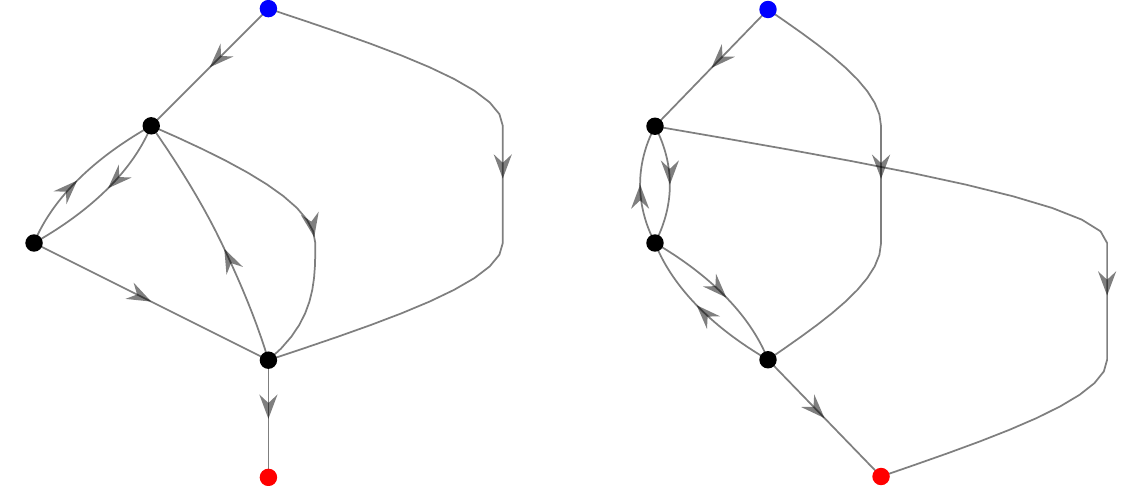}% Here is how to import pix
\caption{ \label{fig:2FGprogenitor5} There are 2 5-vertex 2FG progenitors at $({\color{blue}a},{\color{red}z})$ with $\tilde \beta_2 > 0$. The Betti numbers on the left and right are $(0,0,1,0,\dots)$ and $(0,1,1,0,\dots)$, respectively. Note that removing the target arc from the 2FG progenitor on the left yields the unique 4-vertex 2FG progenitor with $\tilde \beta_2 > 0$.
} 
\end{figure}

\begin{figure}[htbp]
\includegraphics[trim = 0mm 0mm 0mm 0mm, clip, width=\columnwidth,keepaspectratio]{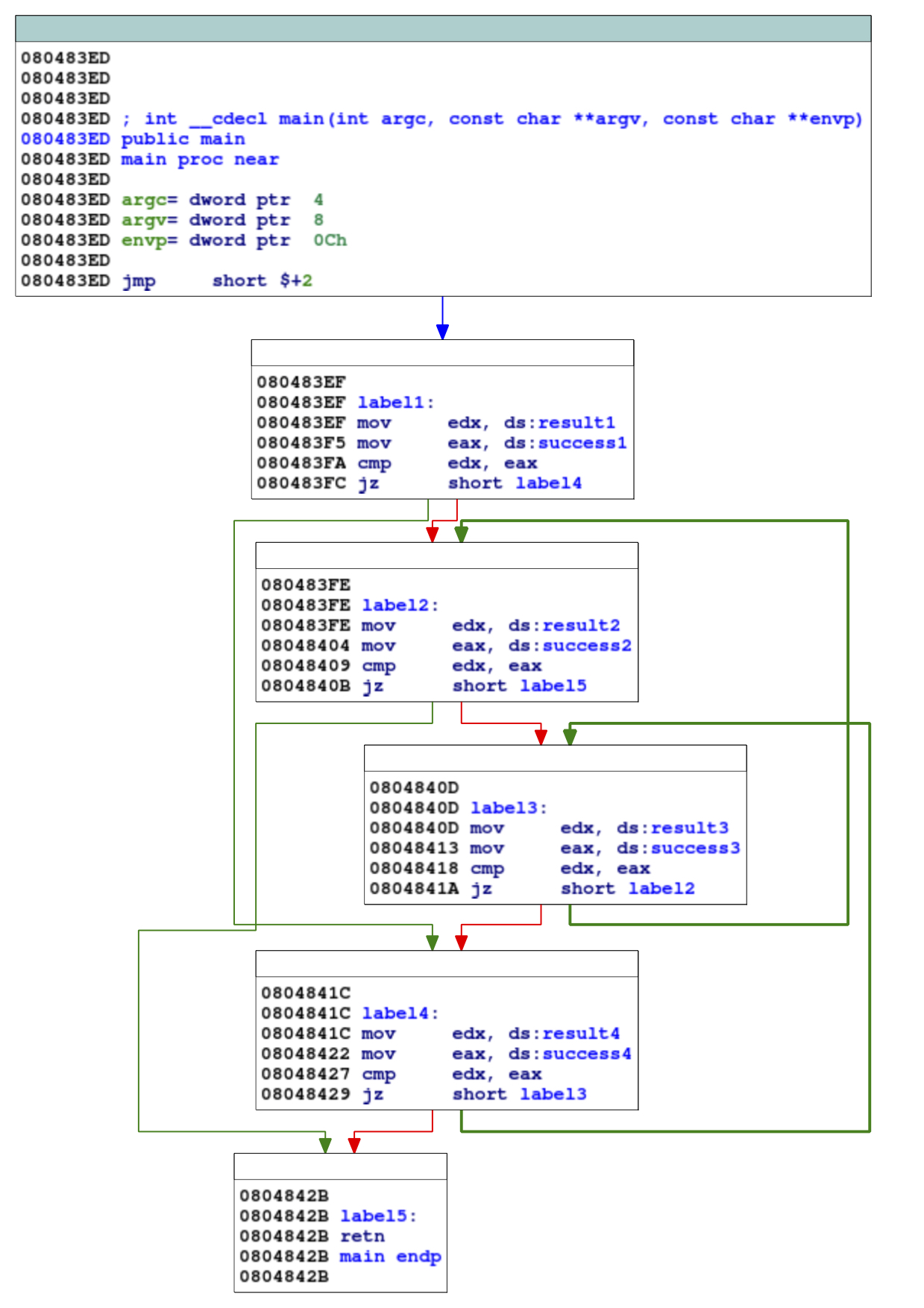}% Here is how to import pix
\caption{ \label{fig:intel_cfg} A control flow example with the (moral) structure of the digraph on the right in Figure \ref{fig:2FGprogenitor5}, shown in IDA Pro \cite{Eagle2011}. The Intel assembly instructions are directly compiled from C code (albeit using \texttt{goto}s and assembly statements). The common instruction motif in all of the basic blocks (i.e., binary code corresponding to vertices) except for the function exit clearly indicates how to construct binaries with control flow given by an arbitrary 2FG (progenitor). %Note that inserting operations without control flow (e.g., arithmetic operations in the instruction set) and reindexing memory addresses at various points would leave the control flow unaffected.
} 
\end{figure}

\begin{figure}[htbp]
\includegraphics[trim = 45mm 95mm 45mm 95mm, clip, width=\columnwidth,keepaspectratio]{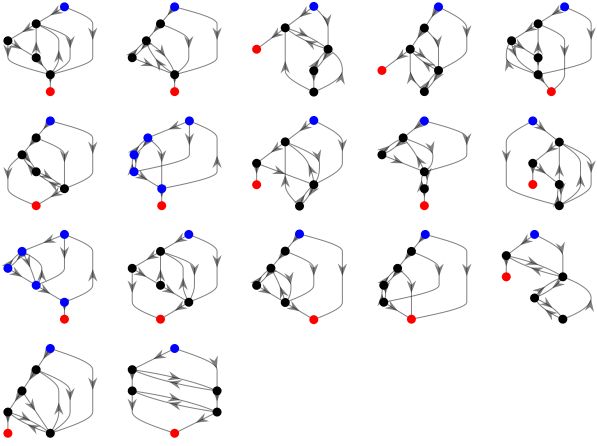}% Here is how to import pix
\caption{ \label{fig:2FGprogenitor6} All 6-vertex 2FG progenitors at $({\color{blue}a},{\color{red}z})$ with $\tilde \beta_2 > 0$. In each case $\tilde \beta_2 = 1$. From left to right and top to bottom, we have $\tilde \beta_1 = 0,0,0,0,1, 1,2,0,0,0, 1,1,1,2,0, 0,0$; all other Betti numbers are zero.
} 
\end{figure}

In Figs. \ref{fig:2FGprogenitor5} and \ref{fig:2FGprogenitor6} we show the 2FG progenitors with $\tilde \beta_2 > 0$ on 5 and 6 vertices, respectively; Figure \ref{fig:intel_cfg} shows an example of disassembled binary code whose control flow is that of the digraph on the right of Figure \ref{fig:2FGprogenitor5}. These examples make it clear that even in a very restrictive setting, the path homology of many different control flow graphs can be nontrivial in dimension $>1$. 

% Untitled20190508

\subsection{\label{sec:Skeletons}Control flow skeletons}

We generated 20000 ``program skeletons'' each resulting from 20 uniformly random productions at uniformly random nonterminals from a context-free grammar along the lines of
\begin{align}
\label{eq:skeleton}
S \rightarrow & S; S \nonumber \\
S \rightarrow & \texttt{if b}; S; \texttt{endif} \nonumber \\
S \rightarrow & \texttt{do while b}; S; \texttt{enddo} \nonumber \\
S \rightarrow & \texttt{repeat}; S; \texttt{until b}
\end{align}
where $;$ is shorthand for a newline. We then analyzed the resulting control flow graphs, in which every line had its own vertex.
%(for an example, see Figure \ref{fig:skeleton}). 
In every case, the equalities $\nu = |\texttt{b}| = \tilde \beta_1$ were satisfied, where here $|\texttt{b}|$ indicates the number of ``predicates'' in the skeleton.
\footnote{
The first equality here is generic, and can be proved via an easy counting argument, but the second equality is not obvious.
} 
Moreover, in every case we also had $\tilde \beta_2 = 0$. This suggests that, e.g., postprocessing, alternative conventions for control flow graph construction, or additional constructs such as \texttt{case} statements are necessary at the source code level in order to obtain nontrivial invariants for structured code in dimension $>1$.

\begin{figure}[htbp]
\centering
\includegraphics[trim = 70mm 110mm 60mm 95mm, clip, width=.5\columnwidth,keepaspectratio]{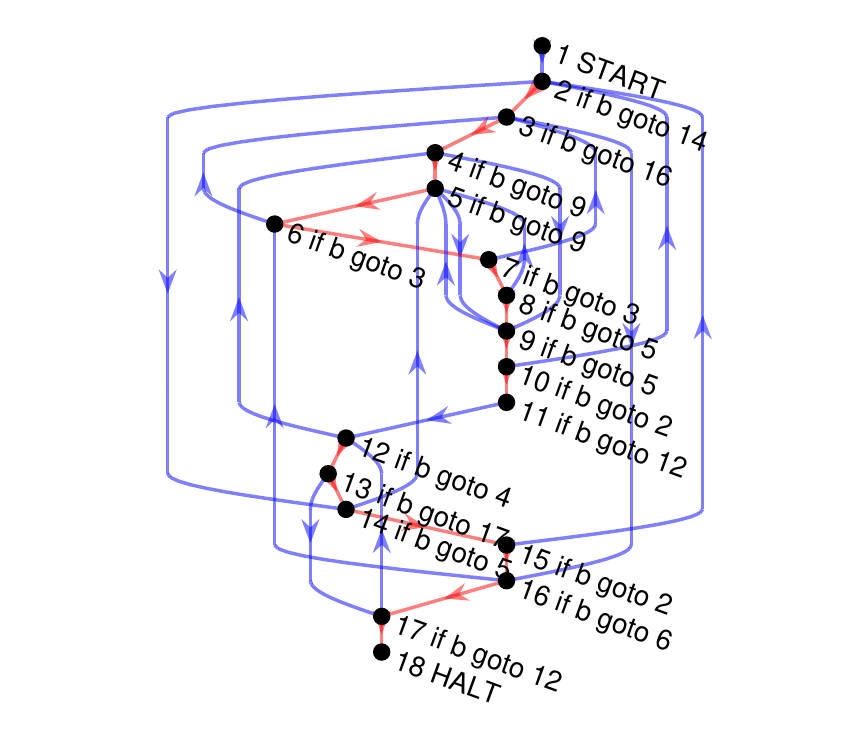}% Here is how to import pix
\includegraphics[trim = 65mm 110mm 65mm 95mm, clip, width=.5\columnwidth,keepaspectratio]{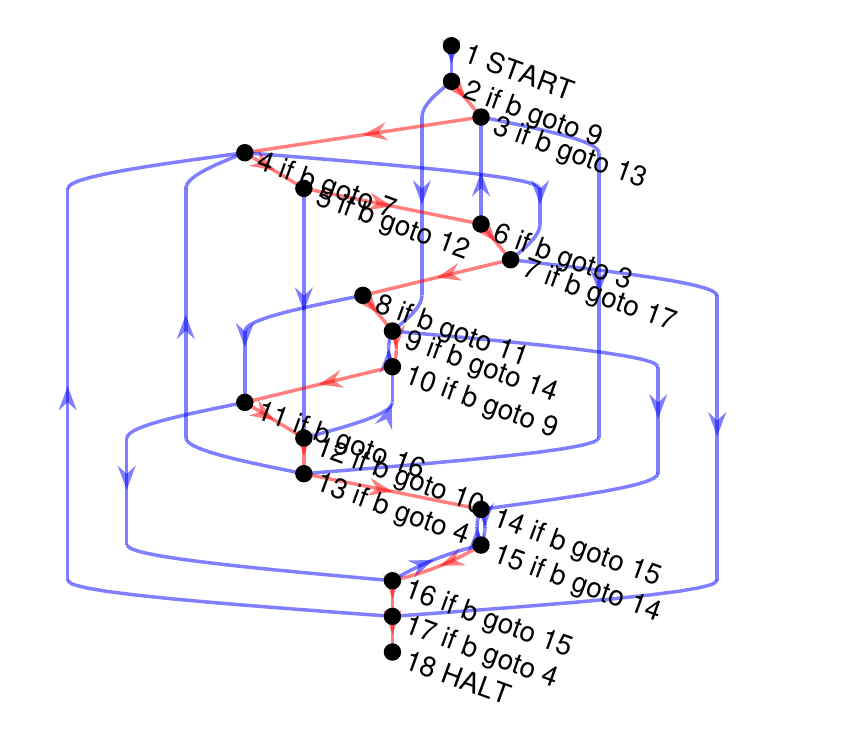}% Here is how to import pix
\caption{ \label{fig:GOTOskeleton} (L) The first of 55 (out of 20000) realizations of a control flow graph for a program skeleton generated through 16 uniformly random conditional \texttt{goto}s that result in $\tilde \beta_2 > 0$. {\color{blue}Blue} (resp., {\color{red}red}) arcs indicate branches where a Boolean predicate (placeholder) \texttt{b} evaluates to {\color{blue}$\top$} (resp., {\color{red}$\bot$}). This particular example has $\tilde \beta_\bullet = (0,11,1,0,\dots)$. (R) The first of the remaining 19945 realizations that \emph{do not} result in $\tilde \beta_2 > 0$. This particular example has $\tilde \beta_\bullet = (0,13,0,\dots)$.
} 
\end{figure}

To illustrate direct applicability to binary code, we also generated 20000 program skeletons whose control flow consisted of 16 conditional \texttt{goto}s to different addresses chosen uniformly at random. In this experiment, 55/20000 skeletons had $\tilde \beta_2 > 0$; we show the first such in the left panel of Figure \ref{fig:GOTOskeleton}. 
%Note the qualitatively different character of this example versus that of Figure \ref{fig:skeleton}. 
In fact, this example is also more complex in other ways than the first skeleton with $\tilde \beta_2 = 0$, shown in the right panel of Figure \ref{fig:GOTOskeleton}: e.g., the digraph drawing in the left panel of Figure \ref{fig:GOTOskeleton} has 15 arc crossings, while the digraph drawing in the right panel has only 10.
\footnote{
Unfortunately, computing the minimal number of arc crossings is NP-hard, so we did not attempt to analyze arc crossings in any more detail.
}

It might seem at first glance that 55 of 20000 control flow graphs is a small enough number to be ignored. However, our experience applying path homology to digraphs obtained from various real-world sources is largely the same as for an exhaustive study of small digraphs of various sorts (to be reported elsewhere, but see various figures throughout the present paper). Namely, subgraphs that generate nontrivial path homology in higher dimensions typically enjoy a high degree of internal structure that is itself a marker of complexity and likely design (or strong compilation) choices, and these are of intrinsic interest for software metrics.

\subsection{\label{sec:Category}Compositional properties and prospects}

If we identify the exit of one flow graph with the entry of another, the resulting Betti numbers are well behaved, adding in the natural way. Specifically, by Proposition 3.25 and Theorem 5.7(a,d) of \cite{Grigoryan2012} (which holds in the non-regular context) along with duality (roughly, $H_p(P \sqcup P') \cong H_p(P) \oplus H_p(P')$), we have the following

\textsc{Proposition.} For flow graphs $F_1$ and $F_2$, let $F_1 \boxtimes F_2$ be the flow graph obtained by identifying the exit edge of $F_1$ and the entry edge of $F_2$. Then 
\begin{equation}
\label{eq:series}
\tilde \beta_\bullet(F_1 \boxtimes F_2) = \tilde \beta_\bullet(F_1) + \tilde \beta_\bullet(F_2). \quad \Box
\end{equation}

While this ``series'' identity is convenient, there are no such straightforward identities for the analogous notions of parallel composition of flow graphs or of inserting one flow graph into another. As a practical matter, it therefore makes sense to analyze complicated flow graphs in a hierarchical and modular fashion using the program structure tree along the lines of \cite{Huntsman2019}: this is not only scalable, but also extends the analogy of cyclomatic complexity in relation to essential complexity \cite{McCabe1976}. 
\footnote{
	Tracking the birth and death of homology classes along these lines via tree persistence \cite{Chambers2018} is a particularly tempting prospect.
}

\section{\label{sec:Wild}Evaluation on disassembled binaries}

\begin{figure}[htbp]
	\centering
	\includegraphics[trim = 50mm 110mm 60mm 115mm, clip, width=.8\columnwidth,keepaspectratio]{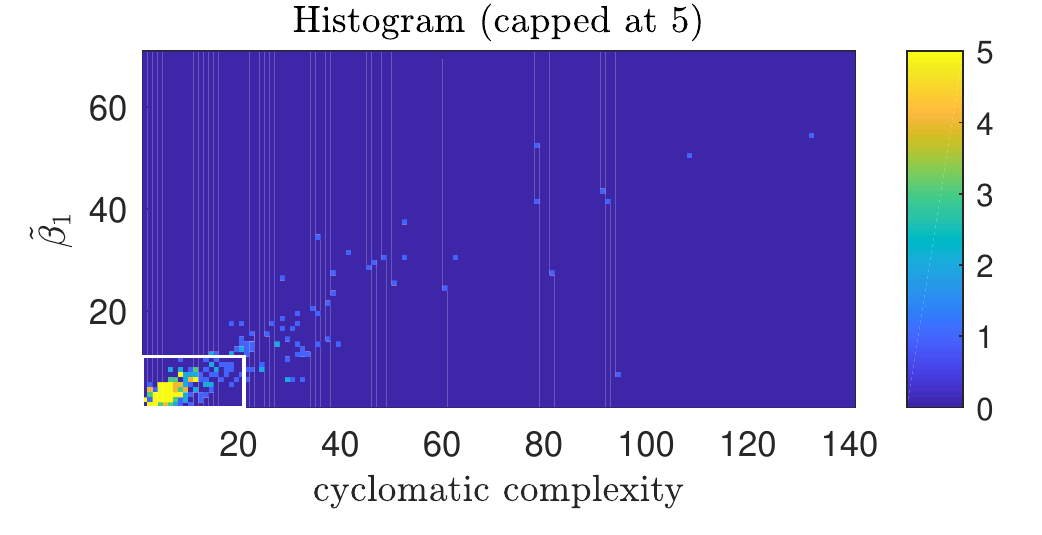}% Here is how to import pix
	\\
	\includegraphics[trim = 50mm 110mm 60mm 115mm, clip, width=.8\columnwidth,keepaspectratio]{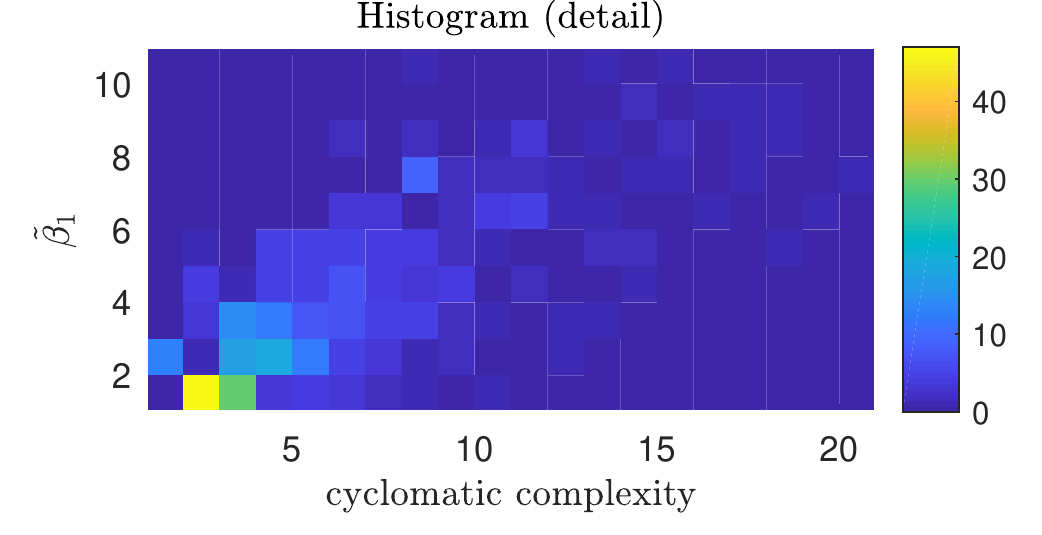}% Here is how to import pix
	\caption{ \label{fig:histogram} (Top) Histogram of $528$ pairs $(\nu,\tilde \beta_1)$ from a variant of \texttt{grep}, where $\nu$ denotes cyclomatic complexity. The color axis is capped at 5. A detail is shown in the bottom panel corresponding to the region in the white box. Besides an outlier at $(93,6)$, the control flow graph corresponding to the vulnerable \texttt{dfamust} function is at $(77,40)$. (Bottom) Detail of the same histogram, without a cap on the color axis.
	} 
\end{figure}

\begin{figure}[htbp]
	\centering
	\includegraphics[trim = 65mm 105mm 60mm 100mm, clip, width=.8\columnwidth,keepaspectratio]{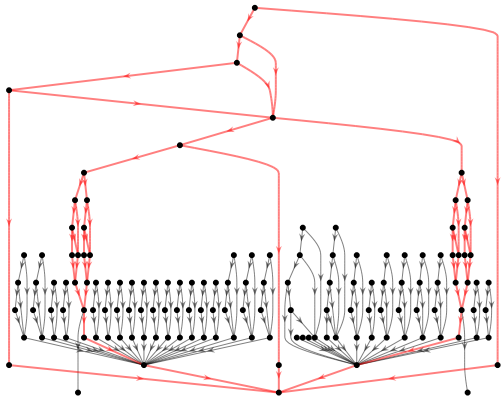}% Here is how to import pix
	\caption{ \label{fig:outlier} Control flow graph corresponding to $(\nu,\tilde \beta_1) = (93,6)$ from Figure \ref{fig:histogram}, with arcs participating in a basis of $\tilde H_1$ highlighted in {\color{red}red}. Not shown: most control flow graphs in this example would have many if not most arcs highlighted in this way.
	} 
\end{figure}

In addition to the preceding analyses, we also considered control flow graphs obtained by disassembling binaries of three variants of the \texttt{GNU grep v2.14} utility with and without injected vulnerabilities (in the \texttt{dfamust} function).
\footnote{The \texttt{grep} code is available as of this writing from the NIST Software Assurance Reference Dataset at \url{https://samate.nist.gov/SARD/testsuite.php\#applications}.}
\footnote{NB. These control flow graphs are generally not flow graphs in the sense of \S \ref{sec:FlowGraphs}.}
Figure \ref{fig:histogram} shows a comparison of the cyclomatic complexity $\nu$ with $\tilde \beta_1$, and Figure \ref{fig:outlier} shows a control flow graph with the anomalous values $(\nu,\tilde \beta_1) = (93,6)$. Although none of the control flow graphs considered here had $\tilde \beta_2 > 0$, Figure \ref{fig:outlier} illustrates how $\tilde \beta_1$ (and the concomitant representation of $\tilde H_1$) can still highlight the most structurally relevant features in a control flow graph.

\section{\label{sec:Conclusion}Conclusion}

Path homology offers several improvements on cyclomatic complexity. With \texttt{case} statements, the Betti numbers can take on arbitrary values for control flow graphs at the source code level. The Betti numbers can also take on nontrivial values in arbitrary dimension for control flow graphs at the assembly level. 

In dimension 1 path homology appears to give the same results as cyclomatic complexity for structured control flow, and arguably more relevant results for generic control flow. While it is obviously desirable to turn evidence for the claim regarding structured control flow into proof, it is already apparent that path homology has the potential to substantially improve upon an archetypal software metric.
%%
%% The acknowledgments section is defined using the "acks" environment
%% (and NOT an unnumbered section). This ensures the proper
%% identification of the section in the article metadata, and the
%% consistent spelling of the heading.
\begin{acks}
We thank Samir Chowdhury, Michael Robinson, and Mikael Vejdemo-Johansson for many conversations regarding path homology, Greg Sadosuk for writing, compiling, and disassembling C code to produce Figure \ref{fig:intel_cfg}, Denley Lam for providing control flow of disassembled \texttt{grep} binaries, and Matvey Yutin for a separate analysis of analogues of various results from \cite{Grigoryan2012} in the non-regular context. This material is based upon work partially supported by the Defense Advanced Research Projects Agency (DARPA) SafeDocs program under contract HR001119C0072. Any opinions, findings and conclusions or recommendations expressed in this material are those of the author(s) and do not necessarily reflect the views of DARPA.
\end{acks}

%%
%% The next two lines define the bibliography style to be used, and
%% the bibliography file.
\bibliographystyle{ACM-Reference-Format}
\bibliography{BAEbib}

%%
%% If your work has an appendix, this is the place to put it.
%\appendix

\end{document}